\documentclass[prb,twocolumn,showpacs,superscriptaddress,groupedaddress]{revtex4-1}

\usepackage{amssymb}
\usepackage{graphicx}
\usepackage{natbib}
\usepackage{epstopdf}
\usepackage{dcolumn}
\usepackage{bm}
\usepackage{makeidx}
\usepackage{enumerate}

\usepackage {graphicx}
\usepackage {amsmath}
\usepackage {amsfonts}
\usepackage {amssymb}
\usepackage {mathrsfs}
\usepackage {bbm}
\usepackage{subfigure}

\newcommand {\be}{\begin{eqnarray}}
\newcommand {\ee}{\end{eqnarray}}

\begin{document}

\title {Robust accidental nodes and zeroes and critical quasiparticle scaling
in iron-based multiband superconductors}
\author {Valentin Stanev}
\affiliation {Materials Science Division, Argonne National Laboratory, 9700 South Cass Avenue, Argonne , IL 60439}
\author {Boian S. Alexandrov}
\affiliation {Theoretical Division, Los Alamos National 
Laboratory, Los Alamos, New Mexico 87545}
\author {Predrag Nikoli\' c}
\affiliation {Department of Physics \& Astronomy, George Mason University, MS 3F3, Fairfax, VA 22030}
\author {Zlatko Te\v sanovi{\' c}}
\affiliation {Institute for Quantum Matter and Department 
of Physics \& Astronomy, The Johns Hopkins University, Baltimore, MD 21218}

\date {\today}

\begin{abstract}
We study multigap superconductivity, with 
strong angular variations of one of the gaps, as appropriate for certain iron-based
high-temperature superconductors. We solve the gap equations 
of this model and find that the nodes or zeroes in the gap function
present at $T_c$ -- although purely  accidental -- typically survive down to $T=0$. 
Based on this result, we investigate the line of quantum 
transitions at which gap zeroes first appear. 
The peculiar "zero-point" critical scaling emanating from this line
dominates quasiparticle thermodynamics and transport properties over much 
of the phase diagram, and supplants more familiar forms of scaling associated 
with the accidental nodes.
\end{abstract}
\maketitle


\section{Introduction}
The discovery of iron-based
high-temperature superconducting family \cite{LaOFeAs} has altered
the landscape of condensed matter research. The underlying physics 
of iron-pnictides and the 
first high-T$_c$ family -- the cuprates -- appear to be 
significantly different. Thus, after two decades 
of cuprate domination, the superconductivity research 
faces new major challenges and new important problems.

The progress in our understanding of iron-pnictides
has been swift \cite{physicaC, Paglione}, but many 
important questions remain unanswered. 
In particular, 
the form of the order parameter remains uncertain. There is growing consensus that the it belongs to the general class of the so-called $s\pm$, $s'$ or extended $s$-wave state \cite{Mazin}, 
the basic dynamical origin of which can be understood from analytic renormalization
group (RG) arguments \cite{Chubukov,Vlad}. This superconducting state is generated by the  electron-electron interaction, originating in Coulomb repulsion. This mandates a sign change between the gaps on the  multiple sheets of the Fermi surface (FS) (interestingly, a similar order parameter appears within the strong
coupling approach to pnictides \cite{Seo}).  
In 122s - the best studied compounds of this family - the 
angle resolved photoemission spectroscopy (ARPES) \cite{Wray, Ding, Zhang} 
detects up to four relatively isotropic gaps, 
one for each of the two hole and two electron bands of the FS. 
This is in an apparent contradiction to 
the penetration depth \cite{lambda, Matsuda}, thermal conductivity \cite{Ames} and specific heat data\cite{Gofryk, Gofryk2}, which indicate
gapless quasiparticle excitations -- a natural
interpretation is that these are associated
with nodes in the gap somewhere along the 
(multiply-connected) FS.  Moreover, Refs. \onlinecite{Ames} and \onlinecite{Gofryk2} report
such
excitations both in the under- and 
over-doped phases of (Co,Ba)122, but they apparently disappear at optimal doping. 
This suggests a significant change in the gap structure as a function of doping. Similar doping evolution of the gap has been reported based on penetration depth measurements \cite{Ames2}. 

Such reading of the experiments receives 
some support from theory. Random phase approximation (RPA)-based calculations \cite{Kuroki,Graser} 
hint at a multitude of competing
nodal and nodeless states and possible transitions among them,
tuned by material parameters like pnictogen height or impurity concentration. 
Furthermore, fluctuation exchange \cite{Schmalian} and numerical functional RG (fRG) 
studies \cite{Thomale,Fa Wang, Thomale2} also find significant 
angular variations and possible nodes in one or more of the gaps. 
A similar possibility was discussed in a model of
Ref. \onlinecite{Chubukovii}.
In these studies, the nodal structure is 
typically induced by the strong {\em orbital anisotropy} of effective interactions
{\em within} the extended $s$-wave state. Thus, such nodes are
{\em "accidental,"} in the sense that they are not protected by any
symmetry or topological considerations. Consequently, their presence 
or absence, and location on the FS are all affected by the 
interactions, temperature, impurity scattering, and the like.


Both the numerical fRG and RPA studies, as well as Ref. \onlinecite{Chubukovii}, 
have focused on the {\em linearized} gap equations and the ensuing pairing 
states at $T=T_c$. However, since the gap equations 
form a complicated non-linear system there is no 
guarantee that the gap structure will remain unchanged, 
and that such accidental nodes or zeroes, present at $T=T_c$, 
survive as $T \rightarrow 0$. 

In this paper we consider a model of a two-band superconductor, 
which has a uniform gap on one band (hole) and gap with angular variations on the other (electron) band. The second gap takes
the form $\Delta_0 + \Delta ' \cos{4\theta}$, which obviously  
allows nodes for $\Delta '> \Delta_0$. As stressed above, both 
nodal and nodeless states belong to the native $A_{1g}$ representation 
of the lattice rotation group and the appearance of nodes is a
direct consequence of the orbital anisotropy of the effective 
interactions.  This model emulates the physics
of iron-pnictides and its different versions have been used in
that context \cite{Chubukovii,Boyd, OptCond, Vekhter}.
A summary of our results is as follows: i) The gap structure obtained at $T=T_c$ 
is fairly robust as a function of $T$ and, in particular, if nodes exist 
near $T_c$ they can survive as $T\rightarrow 0$; ii) at the zero-points $\Delta ' = \Delta_0$, there are gapless quasiparticle excitations with unusual anisotropic dispersion. This leads to a distinctive power-law temperature behavior of many thermodynamic quantities; iii) At $T=0$ these zero-points form a surface of quantum phase transitions with novel and peculiar anisotropic scaling. Consequently, the familiar
Simon-Lee scaling \cite{SL} of 
Dirac-like nodes is superseded by a different form of scaling -- this provides
us a with a new diagnostic tool to apply to the phenomenology
of pnictides. 

\section{The structure of the gap function}
The two-band model reflects the basic features of
iron-pnictides: a hole-like and an electron-like band at the center and at
the corner of a square Brillouin zone (BZ), respectively.
These bands are coupled via interband pairing
interaction $ N(0) V(k, k') =\lambda_0 + \lambda_n \cos{4 \theta}$, where 
$N(0)$ is the density of states (DOS) at the FS, 
$\lambda_0$ and $\lambda_n$ are coupling constants and 
$\theta$ is an angle on the electron FS (using 
coordinate system centered at the electron pocket).
 Defining
gap functions $\Delta_h (\theta)\equiv \Delta_h $ and 
$\Delta_e (\theta) \equiv \Delta_0 + \Delta ' \cos{4\theta}$ 
and introducing intraband repulsion $\mu$ results in 
the gap equations:
\be
\Delta_h & = & -   \int_{0}^{\Lambda}d\xi \int_{0}^{2 \pi}  \frac{d\theta}{2 \pi} \frac{ \tanh{(E_e/2T)}}{E_e}(\lambda_0 + \lambda_n \cos{4\theta}) \Delta_e  \nonumber\\ 
& & -\mu \int_{0}^{\Lambda}d\xi \frac{ \tanh{(E_h/2T)}}{E_h} \Delta_h  \nonumber \\ 
 \Delta_0 & = & -\mu \int_{0}^{\Lambda}d\xi \int_{0}^{2 \pi} \frac{d\theta}{2 \pi} \frac{ \tanh{(E_e/2T)}}{E_e} \Delta_e \nonumber\\
& &-\lambda_0 \int_{0}^{\Lambda}d\xi \frac{ \tanh{(E_h/2T)}}{E_h} \Delta_h \nonumber\\
\Delta ' & = & - \lambda_n \int_{0}^{\Lambda}d\xi \frac{ \tanh{(E_h/2T)}}{E_h} \Delta_h~~,
\label{qe}
\ee 
where $\Lambda$ is a high-energy cut-off and $E_e \equiv \sqrt{\xi^2 + \Delta_e^2} ;\  E_h \equiv \sqrt{\xi^2 + \Delta_h^2} $. Since all interactions are
repulsive $\langle \Delta_e \rangle_{FS} \Delta_h <0$ is
a necessary condition for non-trivial solutions 
of Eqs. (\ref{qe}) to exist ($\langle\cdots\rangle_{FS}$ 
is the angular average over the electron FS). Such sign-switching state is a
likely prospect for pnictides \cite{physicaC}. For $T\rightarrow T_c$, these equations can be linearized with the following result
(see also  Ref. \onlinecite{Chubukovii}): when $\Delta ' = \Delta_0$, there are four zeroes in $\Delta_e$ at $(p_F, 0)$, $(0, p_F)$, $(-p_F, 0)$ and $(0, -p_F)$ (using coordinate system, rotated by $\pi/4$ with respect to BZ axes, and where $p_F$ is the Fermi momentum for the electron band). The same 
structure is reproduced at other corners of the BZ. It is important
to stress here that, while (\ref{qe}) is a simplified phenomenological model, 
the physics and the results that are the focus of this paper will
{\em remain unchanged} when a more realistic description is employed.

\begin{figure}[!h]
\begin{center}
\includegraphics[width=0.4\textwidth]{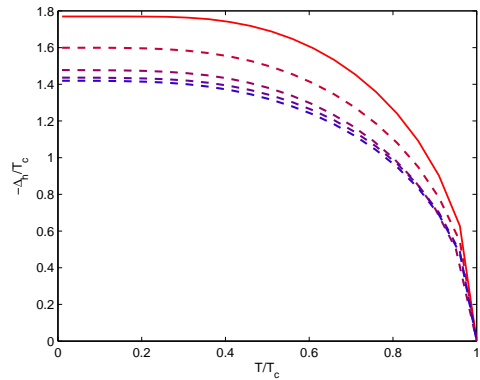}
\end{center}
\caption{$\Delta_h(T)/T_c$ as obtained from solving eqs. \eqref{qe}. The solid red line is for parameter values $\mu=0.05;\lambda_0=0.2$ and $\lambda_n=0$, and corresponds to the BCS result for a single band. Increasing $\lambda_n=0$ from zero leads to \emph {decrease} in $\Delta_h/T_c$ - shown for $\lambda_n=0.1$, $\lambda_n=0.15$, $\lambda_n=0.2$ and $\lambda_n=0.3$ (the blue component increases with increasing $\lambda_n$). Note the negative sign of $\Delta_h$ - consequence of choosing $\Delta_0$ positive.}
\label{fig:1}
\end{figure}

\begin{figure}[!h]
\begin{center}
\includegraphics[width=0.4\textwidth]{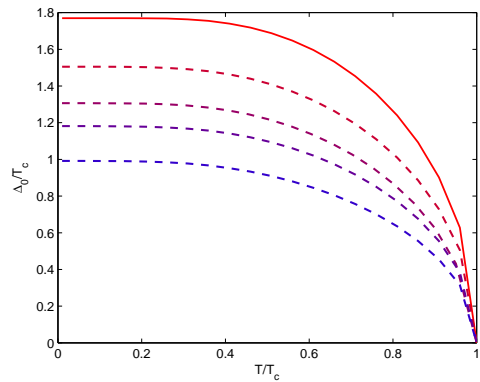}
\end{center}
\caption{$\Delta_0(T)$ obtained from solving eqs. \eqref{qe}. The parameters, color coding and normalization are the same as in Fig. \ref{fig:1}.}
\label{fig:2}
\end{figure} 

\begin{figure}[!h]
\begin{center}
\includegraphics[width=0.4\textwidth]{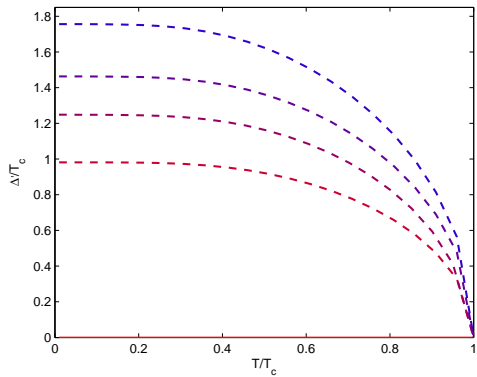}
\end{center}
\caption{$\Delta ' $ as function of temperature. The parameters, color coding and normalization are the same as in  Fig. \ref{fig:1}.}
\label{fig:3}
\end{figure}

\begin{figure}[!h]
\begin{center}
\includegraphics[width=0.45\textwidth]{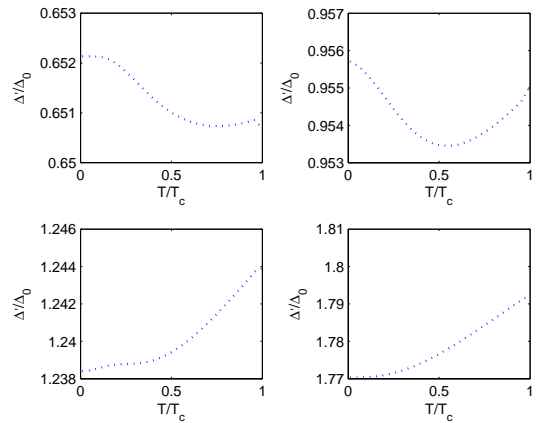}
\end{center}
\caption{Ratio $\Delta '(T)/\Delta '(T)$, arising
from solution of Eqs. \eqref{qe} for $\mu=0.05;\lambda_0=0.2$ and $\lambda_n=0.1, 0.15, 0.2, 0.3$, respectively.  This ratio does not have a simple temperature behavior, but the changes are generically quite small. This means that the nodes, present at $T=T_c$, likely survive as $T\rightarrow 0$.  }
\label{fig:4}
\end{figure}

\begin{figure}[!h]
\begin{center}
\includegraphics[width=0.48\textwidth]{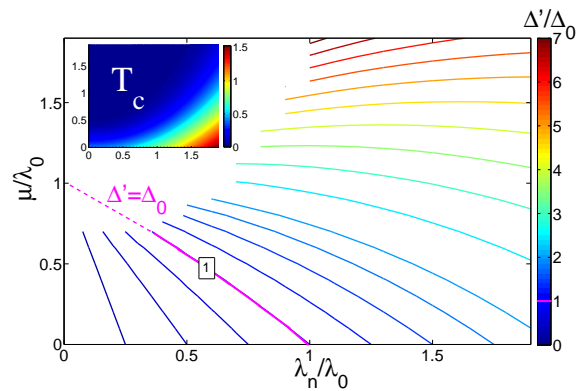}
\end{center}
\caption{The $\Delta '/\Delta_0$ isolines for various $\mu$ and  $\lambda_n$, with $\lambda_0=0.2$. All of them converge at $\mu = \lambda_0$ point at the $y$-axis. The color axis represents $\Delta '/\Delta_0$. For numerical
convenience $T$ is kept finite, but extremely small ($T=10^{-5}T_c$),
and is practically zero. Inset: color map of $T_c$ (in arbitrary units) for the same values of $\mu$ and $\lambda_n$.}
\label{fig:5}
\end{figure}

The first such result follows from solving eqs. \eqref{qe} at arbitrary
$T$, a complicated task which we accomplish numerically.
Representative results for  $\Delta_h$, $\Delta_0$ and $\Delta '$ and are
displayed in Figs. \ref{fig:1}, \ref{fig:2} and \ref{fig:3}. For $\lambda_n=0$ the model reduces to the case of two uniform gaps which follow (for equivalent bands) the BCS temperature dependence (a known result for multiband superconductors \cite{Suhl, Moskalenko}). A nonzero value of $\lambda_n$ generates a finite $\Delta '$, and pushes $\Delta_0/T_c$ and $\Delta '/T_c$ below their BCS values (of course, in absolute units $T_c, \Delta_0$ and $\Delta_h$ all go up).
Evidently, the temperature dependence of different gaps
is {\em similar}. Most importantly, this implies that 
$\Delta '/\Delta _0$ has a rather weak $T$ dependence 
and a nodal state at $T=T_c$ will thus 
likely remain nodal as  $T$ goes to $0$ (shown on Fig. \ref{fig:4}). Remarkably, 
$\Delta '(T)/\Delta_0(T)$ can both increase or decrease 
or can even exhibit a non-monotonic behavior, testifying to the 
strong non-linearity of the gap equation. Nevertheless, the overall changes are
generically quite small. Thus, the $T=T_c$ phase diagram survives at $T=0$ 
without major changes  (Fig. \ref{fig:5} shows our
phase diagram, with $\lambda_0$ fixed). 
We also point out that $\Delta_0(T\rightarrow 0)/\Delta_h(T\rightarrow 0)$ is not equal to $\Delta_0(T\rightarrow T_c)/\Delta_h(T\rightarrow T_c)$  as expected for isotropic 
two-band superconductor \cite{Kresin}.

The above results are surprising since the nodes are only
accidental and therefore sensitive to disruption by the strongly 
non-linear nature of the gap equations \eqref{qe}. Their robustness can be 
viewed as {\em a posteriori} justification for the use of 
nodal or near nodal states in low $T$ calculations, since previously 
these states were self-consistently obtained only at $T_c$.  Note also that
the three coupling constants of our model lead to a certain
ambiguity: a nodal state can be 
created by large $\lambda_n$ \emph {or} by strong $\mu$, 
since intraband repulsion suppresses the uniform gap
 stronger than the oscillating one. These two distinct mechanisms of
generating nodes -- 
a very anisotropic interband interaction and a strong intraband 
repulsion -- are both present in pnictides, and reflect genuine physics of
these multiband superconductors. Furthermore, because the number of gaps in real materials is (at least) four, there is additional frustration associated with interband pair scattering between the gaps with the same sign. This frustration effect also favors the nodal components \cite{Thomale2}, and should be included 
in a more realistic four or five-band calculations.



\begin{figure}[!h]
\begin{center}
\includegraphics[width=0.7\columnwidth]{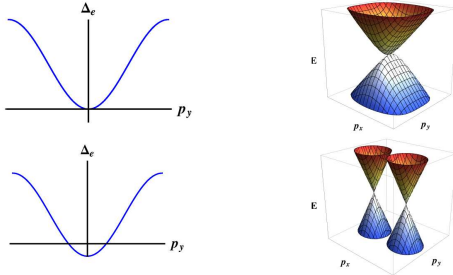}
\end{center}
\caption{The quasiparticle dispersion near accidental
zeroes and nodes. For strongly anisotropic gap,
$|\Delta_0-\Delta'|\ll\Delta_0$, the "zero-point" critical scaling dominates
the nodal contribution.}
\label{fig:6}
\end{figure}

The second key result follows from the first:
consider the surface $\Delta '=\Delta_0$ in 
the ($\lambda_n, \mu, T$) space (with $\lambda_0$ fixed), and ignore
$\Delta_h$ since its contribution is exponentially suppressed at low $T$. 
On this surface the gap in the electron band develops
four zeroes, which, for $\Delta '>\Delta_0$, split into eight nodes. 
At $T=0$, this surface defines a {\em line of quantum phase transitions} 
(see Fig. \ref{fig:5}), along which the low energy quasiparticle spectrum
suffers a non-analytic transformation.
For $\Delta '<\Delta_0$, the quasiparticle excitations are gapped
whereas, for $\Delta '>\Delta_0$, one enters the nodal region, 
with gapless Bogoliubov-deGennes (BdG) quasiparticles 
described by a Dirac-like Hamiltonian
\cite{SL}. At the critical line $\Delta '=\Delta_0$ separating the two regimes --
where the pairs of nodes merge into zeroes -- 
the quasiparticle dispersion has a peculiar character:
linear along the Fermi velocity and quadratic perpendicular to it (see Fig. \ref{fig:6}). 
The resulting spectrum gives rise to an unusual critical 
scaling of low $T$ thermodynamics, 
which dominates the phase diagram surrounding the critical line.

\section{Zeroes in the gap function and their signatures in 
quasiparticle phenomenology}
To explore this peculiar criticality at the $\Delta '=\Delta_0$
line, we expand the gap function in the
BdG Hamiltonian around one of the zeroes, say $( p_F, 0)$: 
\be
\mathcal{H}_{\mathrm{BdG}}= \begin{bmatrix}
\frac{{\bf p}^2}{2m}  - \epsilon_F & \hat{\Delta}({\bf r}) \\
\hat{\Delta}({\bf r})&-\frac{{\bf p}^2}{2m} + \epsilon_F
\end{bmatrix} \approx
\begin{bmatrix}
 v_F p_x& \frac {8 \Delta_0}{ p^2_F} p_y^2\\
\frac {8\Delta_0}{p^2_F} p_y^2&-v_F p_x
\end{bmatrix}~.
\label{bdgH}
\ee
Eq. \eqref{bdgH} results in  an anisotropic dispersion 
$E=\sqrt{(v_F p_x)^2 +(8 \Delta_0 p^2_y/p^2_F )^2}$ 
for quasiparticle energies.
This leads to anisotropic scaling and different 
low-energy ($E\ll\Delta_0$) scaling lengths along $x$ and $y$ directions:
$v_F/E$ and $\sqrt{v_Fa/E}$, respectively, where $a= (8\Delta_0/v_Fp_F^2)$.
In the Lagrangian nomenclature,
the scaling dimensions  at $T=0$ are
$[y] = -1$, $[x] = - z_x$, $[\tau] = - z_{\tau}$,
where $z_\tau = z_x = 2$ \cite{Sachdev}. 

We can now evaluate the DOS of BdG quasiparticles:
\be
N_e(E)= \Bigg\langle \mathrm{Re}\left(\frac{E}{{\sqrt{E^2 - \Delta(\theta)^2}}}\right)\Bigg\rangle_{FS}
\sim \sqrt{E}~
\ee
which is different from the familiar nodal DOS
$N_d(E) \sim E$. This increase of DOS relative to the nodal case is 
caused by the parabolic part in $E$.
 An exact expression, 
in terms of complete elliptic integrals \cite{Maki}, is:
\be
N_e(E)= \frac{N_z}{2 \pi} \sqrt{\frac{E}{\Delta_0}} K(q),\ \mathrm{where}\ \ q=\sqrt{1 +\frac{E}{2 \Delta_0}}~;
\ee
\noindent
where $N_z$ is the number of gap zeroes, and as expected, for $E\rightarrow 0$,
$N_e(E)\rightarrow\sqrt{E}$.

\begin{figure}[!h]
\begin{center}
\includegraphics[width=0.45\textwidth]{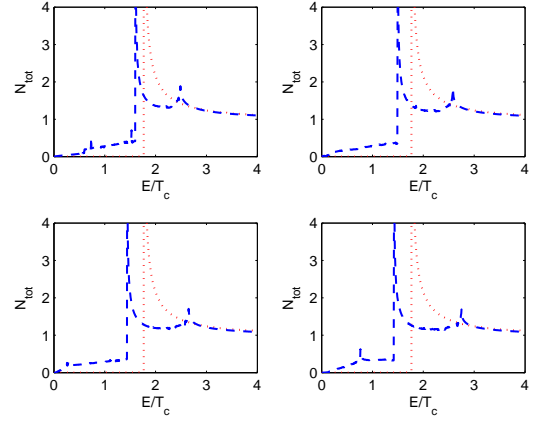}
\end{center}
\caption{The total DOS, calculated using the gap functions for $\lambda_n=0.1, 0.15, 0.2, 0.3$, respectively. In each panel the BCS DOS ($\lambda_n=0$) is included for comparison.}
\label{fig:7}
\end{figure}

We have calculated the total DOS -- $N_{tot}(E,T)=N_{e}(E,T)+N_{h}(E,T)$ -- 
by using the gap functions shown in 
Figs. \ref{fig:1}, \ref{fig:2} and \ref{fig:3}. 
The results (at $T=0$) are displayed in Fig. \ref{fig:7}. 
Clearly, the DOS exhibits a complex behavior with several prominent
 features. First, the two coinciding (at $\lambda_n=0$) BCS singularities 
in $N_e$ and $N_h$ at $E=\Delta_h= \Delta_e$, split for a
nonzero $\lambda_n$. In addition, a new peak appears in 
$N_e$, due to the introduction of a new energy scale $\Delta '$. 
The positions of the three peaks evolve with increasing $\lambda_n$. 
For $\Delta '> \Delta_0$, the low-energy excitations exist all the way down 
to $E=0$, as expected for nodal order parameter.  

Having determined DOS, we are now in position to
extract the low $T$ dependence of various 
thermodynamic quantities. The specific heat can be written as:
\be
C(T) \approx \frac{N_z}{T^2} \int_0^{\infty}dE \frac{E^2 N(E)}{\cosh^2(E/2T)} \sim N_z T^{3/2}.
\label{specheat}
\ee 
Similarly, we can obtain the temperature dependence of the spin susceptibility and the superfluid density: 
\be
 &&\chi_s(T)/\chi_n = 1 - \rho_s(T)/\rho_0 \approx \nonumber\\
&&\approx \frac{N_z}{2 T}\int_0^{\infty}dE  \frac{N(E)}{\cosh^2(E/2T)} \sim N_z T^{1/2},
\ee
from which we can obtain the leading temperature-dependent 
term of the penetration depth:
\be
 \lambda(T) = \sqrt{\frac{c^2 m}{4 \pi e^2\rho_s(T)}}\sim \lambda(0) 
+ \lambda_1 T^{1/2}~.
\ee  
The nuclear spin relaxation rate is:
\be
\frac{(T_1)_n}{(T_1)_s} \approx \frac{N_z}{T}\int_0^{\infty}dE \frac{N(E)^2}{\cosh^2(E/2T)} \sim N_z T.
\label{NMR}
\ee
For the thermal conductivity we need the quasiparticle scattering rate, which can be written as:
\be
\frac{1}{\tau_{{\bf k}}} =N_z\frac{1}{\tau_{n}} \frac{N(E)}{N_n}\left(1-\frac{\Delta_0 \Delta_{\bf k}}{E_{\bf k}^2}\right) \nonumber
\ee 
where $\tau_n^{-1}$ and $N_n$ are the normal state scattering rate and DOS respectively. The thermal conductivity in the superconducting state is: 
\be
\kappa_{ij} \sim \frac{ N_z}{T^2}\int_0^{\infty}dE \frac{E}{\cosh^2(E/2T)} \frac{1}{N(E)} \times \nonumber \\ \int d\Omega\  \hat k_i \hat k_j \left(1-\frac{\Delta_0 \Delta_{\bf k}}{E_{\bf k}^2}\right) \sqrt{E^2-\Delta^2_k}, \nonumber
\ee
where $k_i$ is the $i$th component of the quasiparticle momentum. Integrating around the zeroes and keeping only the term with the lowest power of the temperature, we get for the thermal conductivity, normalized by the normal state value:
\be
\frac{(\kappa_0/T)}{(\kappa_n/T)} \approx \frac{N_z}{T^2}\int_0^{\infty}\frac{dE  E^{5/2}}{N(E) \cosh^2(E/2T)} \sim N_z T.
\label{ThCond}
\ee
 
All of the derived scaling forms strictly apply only at the critical line.
Of course, tuning precisely to a point on this line is 
not an easy task experimentally, and this means that the asymptotic
low temperature behavior of the thermodynamic quantities 
will ultimately be governed either by the 
nodal Dirac or by thermally excited gapped 
BdG  quasiparticles, depending on which side of the line the system finds itself.
Importantly, however, as long as one is 
in the vicinity of the quantum critical line, this asymptotic
behavior will be restricted to a 
{\em very narrow} low temperature range,
 $T\ll |\Delta_0-\Delta '|$, and there will be a
{\em large wedge-shaped crossover regime}, dominated by the quantum phase transition, for which the temperature scalings 
are governed by Eqs. \eqref{specheat}--\eqref{ThCond}. 

\begin{figure}[!h]
\begin{center}
\includegraphics[width=0.4\textwidth]{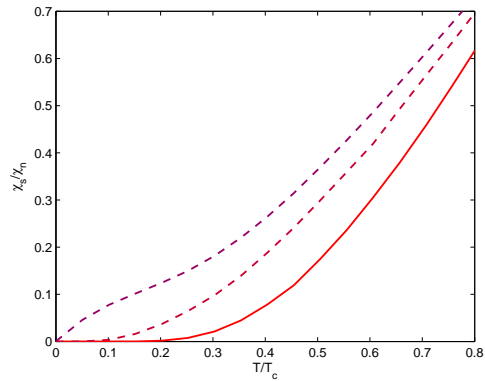}
\end{center}
\caption{The spin susceptibility, calculated using the temperature dependence of the gap functions for $\lambda_n=0, 0.1,$ and $0.15$. The enhancement of the low-temperature part is clearly visible. The case $\lambda_n=0.15$ lies almost on the critical line and thus $\chi_s(T)\sim T^{1/2}$ is expected as a crossover behavior at low temperatures. }
\label{fig:8}
\end{figure}
We supply an illustration of the above ideas by
calculating the spin susceptibility as 
$\Delta '\rightarrow \Delta_0$, and the result is shown in Fig. \ref{fig:8}. 
The low-temperature part of $\chi_s(T)$ is clearly enhanced, and close to $\Delta '= \Delta_0$  (for $\lambda_n=0.15$)  the
$T^{1/2}$ crossover behavior is expected.   

\section{Gap zeroes in finite magnetic field}
Next, we turn to
the problem of a superconductor with gap zeroes in an external magnetic field $H$.
Finite $H$ modifies the critical line in 
the $(\lambda_n, \mu)$ plane, but, for $H$ far below
the upper critical field $H_{c2}(0)$, such modifications are negligible.
The presence of gapless fermions makes the finite $H$ problem 
rather nontrivial, due to the singular scattering 
of fermions from vortices \cite{FT}. Useful results, however, 
can be extracted from general scaling arguments; in nodal $d$-wave 
superconductors such arguments are behind the well-known 
Simon-Lee scaling  \cite{SL}. 


We concentrate on intermediate fields ($H_{c1}\ll H\ll H_{c2}$), 
where the vortex spacing is large and we can assume a uniform field and a
constant gap amplitude. Near the $(p_F, 0)$ zero,
the continuum limit of the gauge-invariant gap function takes the form
$\hat{\Delta}({\bf r}) =(8/p_F^2)\{\partial_y,\{\partial_y,\Delta ({\bf r})\}\} 
+ (2 i/p_F^2)\Delta({\bf r}) \partial_y^2 \phi$, 
where $\Delta({\bf r})=|\Delta| e^{i \phi ({\bf r})}$ and $\{,\}$ 
signifies full symmetrization \cite{FT}. This allows us to 
write the low-energy BdG Hamiltonian as:
\be
\mathcal{H}_{\mathrm{BdG}}=
\begin{bmatrix}
\frac{1}{2m}(\hat{{\bf p}} +\frac{e}{c}{\bf A})^2  - \epsilon_F & \hat{\Delta}({\bf r}) \\
\hat{\Delta}^*({\bf r})&-\frac{1}{2m}(\hat{{\bf p}} - \frac{e}{c}{\bf A})^2  + \epsilon_F
\end{bmatrix}\nonumber
\ee
\be
\approx
\begin{bmatrix}
v_F(p_x+\frac{e}{c}A_x)& \hat{\Delta}({\bf r})\\
\hat{\Delta}^*({\bf r})&-v_F(p_x-\frac{e}{c}A_x)
\end{bmatrix}~.
\label{HamlMF}
\ee

We now choose Landau gauge $A_x = H y$ and rewrite
$v_F(p_x + (e/c) H y)$ in dimensionless form by introducing a
length $d$ and defining variables $x'=x/d$ 
and $y'=yd/l^2$, where $l=\sqrt{c/eH}$ is the magnetic length. 
The off-diagonal terms in $\mathcal{H}_{\mathrm{BdG}}$ are similarly rewritten as
$(8\Delta_0/p_F^2)[\{\partial_y,\{\partial_y,e^{i \phi}\}\} 
+ (i/4) e^{i \phi}  (\partial_y^2 \phi) ] 
= (v_F a d^2/l^4)[\{\partial_{y'},\{\partial_{y'},e^{i \phi}\}\} 
+ (i/4) e^{i \phi}  (\partial_{y'}^{2} \phi)]$ and its complex conjugate. 
The choice $d=(l^4/a)^{1/3}$ allows us to reexpress 
the quasiparticle energy spectrum as
$\{E_n\}\to (v_F/d) \{\varepsilon_n\}=v_F(a/l^4)^{1/3}\{\varepsilon_n\}
= v_F (ae^2H^2/c^2)^{1/3} \{\varepsilon_n\}$, 
where $\{\varepsilon_n\}$ are numbers which in principle can
depend only on $H/(c/ea^2)\equiv H/H_a
\ll 1$.
If they do, such ``anomalous" scaling
would not be surprising in our case of singular anisotropy. 
Adding the contributions from all zeroes 
restores the $x \leftrightarrow y$ symmetry. 

Using the above rescaling, we write DOS $N(E,H)$ as:
\be
\sim{\sqrt{v_F}}(ae^2H^2/c^2)^{1/6}
{\cal N}(\frac{E}{v_F (ae^2H^2/c^2)^{1/3}};\frac{H}{H_a})~,
\label{DOSH}
\ee
where ${\cal N}(u;w)$ is the scaling function.
If $H$ induces finite DOS at low $E$, we can follow earlier
discussion to obtain $N(E\to 0,H)\sim H^{1/3 + \eta}$,
where $\eta > -1/3$ is the "anomalous" dimension, defined
by ${\cal N}(u=0;w\to 0)={\rm const.}\times w^\eta$.
$\eta\not =0$ is clearly a possibility given the fact that
the vortex lattice structure
in $\exp(i\phi ({\bf r}))$ is determined by {\em all} 
electrons and does not conform to the anomalous scaling of
individual zero-points. The precise value of $\eta$ follows from direct 
computation \cite{footbox}.

The critical ``zero-point" scaling also works when both $H$ and $T$ are finite.
The internal energy density can be written as
  \be
U(H, T)& =&  \frac{N_z}{V} \sum_{n}E_n(H)f(E_n(H)/T)\sim \\
  &\sim & N_z H^{5/3}{\cal F}_u(T/(v_F^3 a e^2c^{-2}H^2)^{1/3}; H/H_a)~\nonumber ,
\label{U}
\ee
where $f$ is the Fermi function and
${\cal F}_u$ is a scaling function. Here we
converted $\sum_n\to (V/4\pi^2)\int dp_xdp_y$,
$dp_x dp_y \sim (1/l^2) dp_{x'}dp_{y'} $ \cite{FT}.
 From \eqref{U}, the specific heat per unit volume is:
\be
  C(H, T)  &\sim& N_z H^{5/3}\partial_T {\cal F}_u(T/(v_F^3 a e^2c^{-2}H^2)^{1/3}; H/H_a)\sim \nonumber \\
&\sim&  N_z H {\cal F}_c(T/(v_F^3 a e^2 c^{-2}H^2)^{1/3}; H/H_a)~,
\ee
where ${\cal F}_c(T/(v_F^3 a e^2c^{-2}H^2)^{1/3}; H/H_a)$ is the
corresponding scaling function. 

In general, evaluating ${\cal F}_u$
or ${\cal F}_c$ is a difficult task. However, the needed limits
of these functions are readily deduced: Eq. (\ref{specheat}) mandates
${\cal F}_c( \alpha \rightarrow \infty; w \to 0 ) \rightarrow \alpha^{3/2}$,
where $\alpha = T/(v_F^3 ae^2c^{-2} H^2)^{1/3} $. In the opposite
case, $\alpha \rightarrow 0$, we can use the constant low $E$ DOS, 
induced by $H$,
and observe that (\ref{specheat}) gives $C(H, T) \sim T H^{1/3 + 
\eta}$, and thus
${\cal F}_c(\alpha \rightarrow 0; w\ll 1) \rightarrow \alpha w^{\eta}$.
Note that these scaling arguments apply only
to the critical BdG fermions inhabiting the zeroes. We are not
including the contributions from other parts of the system, like
possible localized states in the vortex cores -- this
contribution is subleading at low $E$, $T$ and $H$.

Again, we want to emphasize that although precise tuning to a state with gap zeroes 
seems very unlikely, close to such state and at not too small $T$ and $H$ the scaling associated with gap zeroes dominates. Thus, in a state with accidental nodes the Simon-Lee scaling can be completely unobservable.   

\section{Conclusions}
In summary, we have studied an anisotropic 
two-band model of iron-based superconductors.
By explicit solution of the gap equations, we uncovered the robust
nature of the accidental nodes and zeroes in the gap function. We have discussed
in some detail the quantum critical line where the gap zeroes first appear. It projects
a considerable influence over the phase diagram and is characterized by a
peculiar form of anisotropic critical scaling, qualitatively distinct 
from the familiar Dirac-Simon-Lee scaling. Irrespective of whether one can
tune in precisely to this quantum critical line in a particular experiment,
as long as the gap is strongly anisotropic 
($|\Delta_0-\Delta'|\ll\Delta_0$),
the "zero-point" critical scaling will dominate the quasiparticle
thermodynamics and transport over a wide region in a 
phase diagram, overwhelming the contribution from accidental nodes.

\section{Acknowledgments}
Work at the Johns Hopkins-Princeton Institute
for Quantum Matter were supported by the U.\ S.\ Department of
Energy, Office of Basic Energy Sciences, Division of Materials Sciences and
Engineering, under Award No.\ DE-FG02-08ER46544. Research at Los Alamos National Laboratory is carried out under the auspices of the U.S. Department of Energy under Contract No. DE-AC52-06NA25396. P.N. is supported by the Office of
Naval Research grant N00014-09-1-1025A.

\bibliographystyle{apsrev}


\end {document}